\documentclass[12pt,letter]{emulateapj081310}

\usepackage{amsmath,amssymb}
\usepackage{xspace}
\usepackage{graphicx}
\usepackage{epstopdf}
\usepackage{graphicx}
\usepackage{epsfig}

\slugcomment{}

\shorttitle{A Compton-thick AGN at $z\sim5$}
\shortauthors{Gilli et al.}

\begin{document}

\title{A Compton-thick AGN at $z\sim5$ in the 4 Ms $Chandra$ Deep Field South}


\author{R. Gilli\altaffilmark{1}, 
J. Su\altaffilmark{2}, 
C. Norman\altaffilmark{2,3}, 
C. Vignali\altaffilmark{4}, 
A. Comastri\altaffilmark{1},
P. Tozzi\altaffilmark{5},
P. Rosati\altaffilmark{6},
M. Stiavelli\altaffilmark{3},
W.N. Brandt\altaffilmark{7,8},
Y.Q. Xue\altaffilmark{7,8},
B. Luo\altaffilmark{7,8},
M. Castellano\altaffilmark{9},
A. Fontana \altaffilmark{9},
F. Fiore \altaffilmark{9},
V. Mainieri \altaffilmark{6},
A. Ptak \altaffilmark{10}
}

\altaffiltext{1}{INAF - Osservatorio Astronomico di Bologna, via Ranzani 1, 40127, Bologna, Italy}
\altaffiltext{2}{Department of Physics and Astronomy, Johns Hopkins University, Baltimore, MD 21218, USA}
\altaffiltext{3}{Space Telescope Science Institute, 3700 San Martin Drive, Baltimore, MD 21218, USA}
\altaffiltext{4}{Dipartimento di Astronomia, Universit\`a degli Studi di Bologna, Via Ranzani 1, 40127 Bologna, Italy}
\altaffiltext{5}{INAF - Osservatorio Astronomico di Trieste, via Tiepolo 11, 34131, Trieste, Italy}
\altaffiltext{6}{European Southern Observatory, Karl Schwarzschild Strasse 2, 85748 Garching bei Muenchen, Germany}
\altaffiltext{7}{Department of Astronomy and Astrophysics, Pennsylvania State University, University Park, PA 16802, USA}
\altaffiltext{8}{Institute for Gravitation and the Cosmos,Pennsylvania State University, University Park, PA 16802, USA}
\altaffiltext{9}{INAF - Osservatorio Astronomico di Roma, via Frascati 33, 00040 Monteporzio Catone, Italy}
\altaffiltext{10}{NASA/GSFC, Greenbelt, MD 20771, USA}
\email{Electronic address: roberto.gilli@oabo.inaf.it}



\begin{abstract}

We report the discovery of a Compton-thick Active Galactic Nucleus (AGN) at $z=4.76$ in the 4 Ms $Chandra$ Deep Field South.
This object was selected as a $V$-band dropout in HST/ACS images and previously recognized as an AGN from optical spectroscopy. The 4 Ms $Chandra$ observations show a significant ($\sim4.2\sigma$) X-ray detection at the $V$-band dropout position. The X-ray
source displays a hardness ratio of HR=$0.23\pm0.24$, which, for a source at $z\sim5$, is highly suggestive of
Compton-thick absorption. The source X-ray spectrum is seen above the background level 
in the energy range of $\sim0.9-4$ keV, i.e., in the rest-frame energy range of $\sim 5-23$ keV. When fixing the photon index to $\Gamma=1.8$, the measured column density is $N_H=1.4^{+0.9}_{-0.5}\times 10^{24}$ cm$^{-2}$, which is Compton-thick. To our knowledge, this is the most distant heavily obscured AGN, confirmed by X-ray spectral analysis, discovered so far. 
%The 0.5-2 keV and 2-10 keV fluxes, as extrapolated from the X-ray spectrum, are
%$4.2\times 10^{-17}$ erg cm$^{-2}$ s$^{-1}$ and $6.8\times 10^{-16}$ erg cm$^{-2}$ s$^{-1}$, respectively. 
The intrinsic (de-absorbed), rest-frame luminosity in the 2-10 keV band is $\sim 2.5\times 10^{44}$ erg s$^{-1}$, which places this object among type-2 quasars.
%Its optical spectrum displays a narrow Ly$\alpha$ emission feature (FWHM$\sim$1000 km s$^{-1}$), consistent with a classification of obscured, type-2 nucleus. 
The Spectral Energy Distribution shows that massive star formation is associated with obscured black hole accretion. This system may have 
then been caught during a major co-eval episode of black hole and stellar mass assembly at early times.
%The diagnostic ratio between the 2-10 keV and 6$\mu$m rest frame luminosity would again 
%suggest that the object is Compton-thick. 
%We stress that heavily obscured, Compton-thick AGNs at very high redshift can only be discovered through very deep and sensitive X-ray observations.
The measure of the number density of heavily obscured AGN at high redshifts will be crucial to reconstruct the BH/galaxy evolution history from the beginning. 

\end{abstract}


%\section{Introduction}
%While optically bright quasars are the most spectacular expression of accretion onto supermassive black holes (SMBHs) at 
%galaxy centers, it is widely believed that SMBHs grow most of their mass during obscured phases, in which
%the detection of the nuclear power becomes challenging \citep[e.g.,][]{fabian99}. 
%\begin{thebibliography}{}
%\bibitem[{{Fabian}(1999)}]{fabian99}
%{Fabian}, A.~C. 1999, \mnras, 308, L39
%\end{thebibliography}

\section{Introduction}

While optically bright quasars are the most spectacular expression of accretion onto supermassive black holes (SMBHs) at 
galaxy centers, it is widely believed that SMBHs grow most of their mass during obscured phases, in which
the detection of the nuclear power becomes challenging \citep[e.g.,][]{fabian99}. 
Large amounts of gas and dust are found to
hide the majority of Active Galactic Nuclei (AGN) in the nearby and distant Universe, as demonstrated by deep and wide
X-ray surveys over different sky fields (see e.g., \citealt{bh05} for a review). From $\sim$30 to $\sim$50\% of all
AGN are believed to be obscured by extreme gas column densities above $N_H=\sigma_T^{-1}\sim10^{24}$cm$^{-2}$. These objects are dubbed ``Compton-thick''
and represent the most elusive members of the AGN population.
The evidence for an abundant population of local Compton-thick objects is compelling: up to $\sim50\%$ of nearby Seyfert 2s contain a Compton-thick nucleus \citep{guido99,akylas09};  about 50 objects - mostly local - have been certified as ``bona-fide'' Compton-thick AGN by X-ray spectral analysis \citep{c04}.

Synthesis models of the X-ray background (XRB)
suggest that Compton-thick AGN must be abundant at least up to $z\sim1$ to explain the peak of the XRB at 30 keV \citep[see e.g.,][and references therein]{gch07,tuv09}. A population of distant, Compton-thick AGN, as abundant as that predicted
by XRB synthesis models, is also required to match the SMBH mass function measured in nearby galaxies with that of
``relic'' SMBHs grown by accretion \citep[e.g.,][]{marconi04}. In recent years it has been proposed that Compton-thick AGN 
represent a key phase of the BH/galaxy coevolution, during which the BH is producing most of its feedback into the host galaxy \citep[e.g.,][]{daddi07,menci08},
and it has also been suggested that their number density steeply increases with redshift \citep{tuv09}.

The observation of heavily obscured AGN at high-$z$, $z\gtrsim2-3$, remains challenging and it is very difficult to estimate their abundance since they produce only a small fraction of the XRB emission and are thus poorly constrained by synthesis models.
%An estimate of how abundant heavily obscured AGN are at high-z, i.e., $z\gtrsim2-3$, is however very difficult. On one hand, high-z
%Compton-thick AGN produce only a small fraction of the XRB emission, and are therefore poorly constrained by synthesis models. On the other hand,
%the combination of obscuration and dimming with distance makes the detection of such a population extremely challenging.
Deep X-ray surveys have proven effective in revealing a few ``bona fide''  Compton-thick AGN at high-z. For instance, four such objects
at $1.53<z<3.70$ have been discovered in the $Chandra/XMM$ - Deep field South \citep[CDFS; see][]{norman02,comastri11,feruglio11}. 
Other examples of candidate Compton-thick AGN at high-z have been reported \citep[e.g.,][]{tozzi06,polletta08}, even up to z=5.8 \citep{brandt01}, albeit with poorer X-ray photon statistics. Selection techniques based on the strength of the mid-IR emission with respect
to the optical and X-ray emission have also been developed and applied to select large populations of candidate Compton-thick AGN up to $z=2-3$
\citep{daddi07,alex08,fiore09,bauer10}. Once more, however, the lack of X-ray spectra prevents an unambiguous
determination of the absorbing column density, making the measurements by these works largely uncertain.

In this paper we report the discovery of a ``bona-fide'' Compton-thick AGN at $z$=4.76 in the 4 Ms CDFS. A concordance cosmology with $H_0=70$~km~s$^{-1}$~Mpc$^{-1}$, 
$\Omega_m=0.27$, $\Omega_{\Lambda}=0.73$ is adopted throughout this paper.

%hiding in an actively star forming, bright sub-mm galaxy \citep{coppin09}.

\section{Source selection and Observations}

We searched for $V-$band dropout objects in the HST/ACS v2.0 data of GOODS-South \citep{java04} associated with X-ray emission
in the 4 Ms $Chandra$ image. We used the $V_{606}$-dropout selection criteria from \citet{oesch07}, which effectively pick sources
at $4.7<z<5.7$\footnote{An object is defined as a  $V_{606}$-dropout if $V_{606}-i_{775}>min[1.5+0.9(i_{775}-z_{850})$, 2], $V_{606}-i_{775}>1.2$, $i_{775}-z_{850}<1.3$, $S/N(z_{850})>5$ and $S/N(B_{435})<3$.}. Details on the production of the $V$-dropout catalog are given in \citet{su11}. Additionally, we required a stellarity parameter (CLASS\_STAR) greater than 0.9 in the $z_{850}$ band to choose point-like sources. This led to 21 star-like $V_{606}$-dropouts with $z_{850}<25.5$, among which there are four z$\sim$5 galaxies, eleven stars, three lower-redshift galaxies, and one z$\sim$5 AGN,  which is the only object detected in X-rays (XID403 in the 4 Ms CDFS catalog of \citealt{xue11}). The remaining two candidates have not been identified spectroscopically.  The measured AB magnitudes of XID403 in ACS images are: $V_{606}=26.84\pm0.10, i_{775}=25.21\pm0.04, z_{850}= 25.05\pm0.04$. The 5$\sigma$ detection limit in the $B_{435}$-band is 28.4 AB mag.

XID403 ($\alpha_{J2000}$=03:32:29.29, $\delta_{J2000}$=-27:56:19.5) was recognized as an AGN at $z=4.76$ based on FORS-2 spectroscopy \citep{vanzella06}. Its optical spectrum exhibits a narrow (FWHM$\lesssim1000$ km s$^{-1}$) Ly$\alpha$ emission line and a broader (FWHM$\sim2000$ km s$^{-1}$) NV$\lambda$1240 emission line, with an integrated flux similar to Ly$\alpha$. A more recent spectrum with DEIMOS/Keck confirms both features \citep{coppin09}.
%We retrieved the one-dimensional FORS-2 spectrum and measured the FWHM of the Ly$\alpha$ line, which is found to be 1000 km/s, suggesting
%a type-2 AGN classification. 
The Spectral Energy Distribution of  XID403 
was published by \citet{coppin09}. Based on a LABOCA detection at 870$\mu$m, they showed that this source is a bright submillimeter galaxy with SFR$\sim$1000 $M_{\odot}$ yr$^{-1}$. A large reservoir of molecular gas ($\sim1.6\times10^{10}\,M_{\odot}$) was also identified through CO(2-1) observations \citep{coppin10}.

We considered the same optical to mid-IR datasets used by \citet{coppin09} and improved on the SED by adding the detection at 1.1 mm by AzTec/ASTE ($f_{1.1mm}=3.3\pm0.5$mJy; Scott et al. 2010) and the $Y$, $J$ and $Ks$ magnitudes from the deep NIR imaging by HAWK-I/VLT ($Y_{AB}=24.56\pm0.12$, $J_{AB}=24.37\pm0.14$, $K_{AB}=24.03\pm0.20$; \citealt{castellano10}). This object is also detected (at $\sim3\sigma$) at 1.4 GHz with a peak flux of 22.3 $\mu$Jy (N. Miller priv. comm).
Unfortunately, it falls just outside the areas covered by the $16\mu$m Spitzer/IRS mosaic \citep{teplitz11} and GOODS-Herschel (PI D. Elbaz). XID403 is not detected in the 3 Ms $XMM$ image of the CDFS. 

%In the near-IR the source went undetected both in J and Ks images taken with the NTT/SOFI instrument down to 23.1 and 22.2 AB mag
%(Olsen et al. 2006), and in J and K images taken with the CTIO (Taylor et al. 09) down to similar limiting magnitudes.
%[Check for a very faint spot at the source location in the H-band NTT/SOFI]. In the mid-IR, the source
%has been detected by Spitzer in all 4 IRAC bands (3.5,4.5,5.8 and 8.0 $\mu$m) and at 24 $\mu$m with MIPS
%(ref). It is not detected at 1.4 GHz down to 30$\mu$Jy. 

\begin{figure*}
\includegraphics[angle=0,scale=.47]{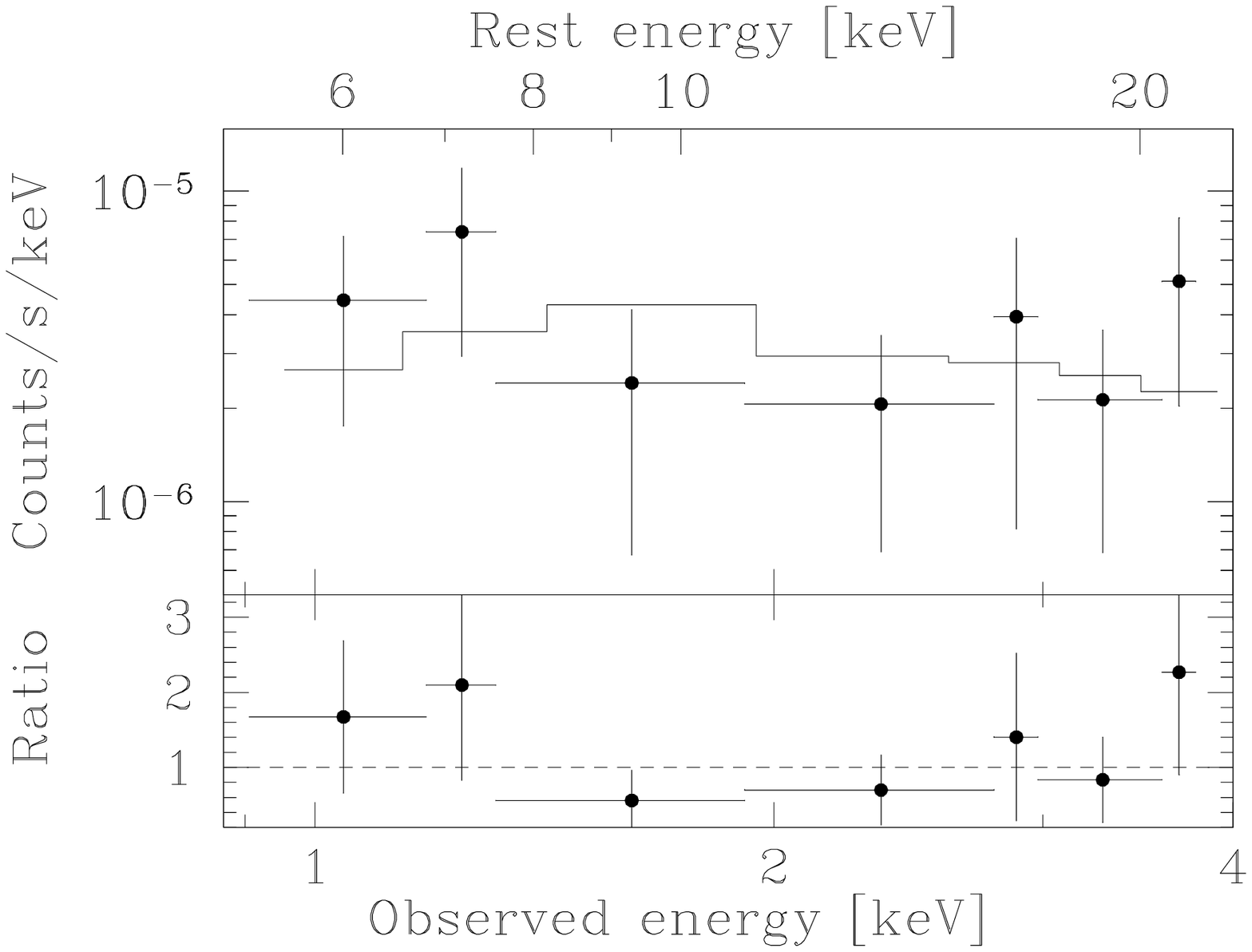}
\hfill
\includegraphics[angle=0,scale=.47]{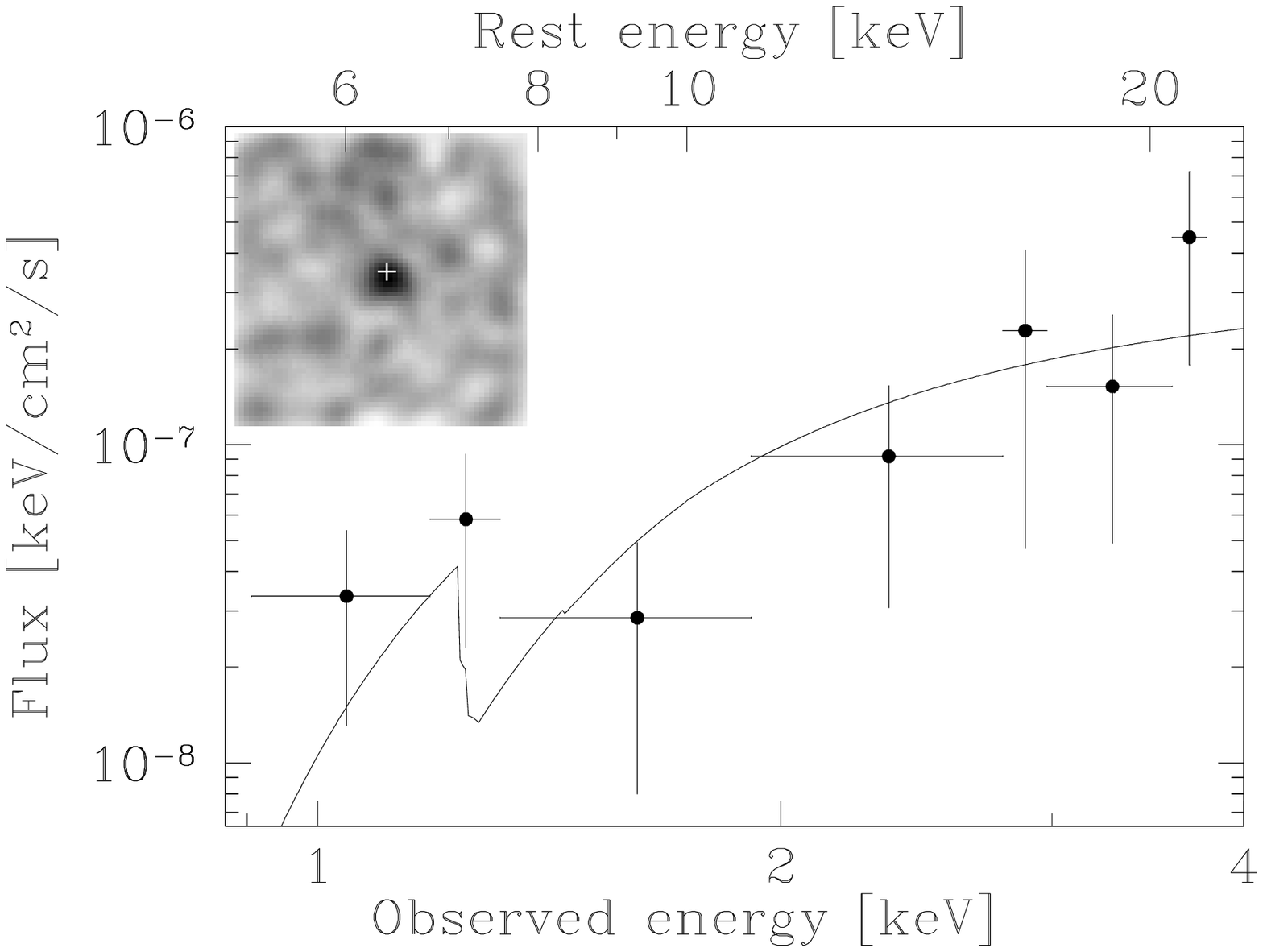}
\caption{ $Left:$ The observed 4 Ms $Chandra$ X-ray spectrum of XID403 and best-fit model (data/model ratio in the bottom panel). A best fit column density of $N_H=1.4^{+0.9}_{-0.5}\times 10^{24}$ cm$^{-2}$ is obtained when assuming $\Gamma=1.8$. The spectrum has been rebinned for display purposes. $Right:$ Response corrected spectrum and best fit model. The inset shows a smoothed 0.9-4 keV image of XID403 (size is 30''$\times$30"). The cross marks the position of the optical $V$-dropout.}
\end{figure*}

\section{X-ray data analysis}

A total exposure of $\lesssim 4$ Ms has been accumulated on the CDFS as a result of 54 individual observations with ACIS-I performed
during three different time periods:  $\sim0.8$ Ms in 2000, $\sim1$ Ms in 2007 and $\sim 2$ Ms in 2010.  X-ray data products, including event files for each observation and also for the merged dataset are publicly available\footnote{ \url{http://cxc.harvard.edu/cda/Contrib/CDFS.html}}. In this paper we use the data products
by \citet{xue11} who derived X-ray source catalogs from a full reprocessing and astrometric recalibration of the event files. 
We used CIAO v4.1 and the Funtools package\footnote{\url{https://www.cfa.harvard.edu/\~john/funtools}} to perform X-ray aperture photometry at the position of XID403.
The separation between the optical and X-ray centroids is $\sim 0.4"$, which is well within the $1\sigma$ X-ray source positional uncertainty ($\sim$0.47'').
To maximize the S/N ratio, we measured the source counts in different bands within a small aperture of 3'' radius, which encloses $\sim$50\% of the PSF
at 1.5 keV at the source location ($\sim8$ arcmin off-axis). 
We measured $37.0\pm 8.7$ net counts in the 0.9-4 keV band, corresponding to a $\sim4.2\sigma$ detection\footnote{XID403 has $82\pm25$ net
counts  in the 0.5-8 keV band (100\% PSF) in \citet{xue11}, in agreement with our estimate.}. We verified that similar results are obtained when using  different local background regions.  The hardness ratio, defined as HR = (H-S)/(H+S),  where S and H are the net counts observed in the 0.5-2 keV and 2-7 keV bands, respectively, is HR=$0.23\pm0.24$ (not corrected for vignetting).  This value, for an AGN at $z\sim5$ with a standard intrinsic spectrum (i.e., $\Gamma=1.8$) is highly suggestive of heavy obscuration. For comparison, an AGN at $z\sim5$ with $N_H\lesssim10^{23}$cm$^{-2}$ is expected to have HR$\lesssim-0.35$ at the source position.
We extracted the X-ray spectrum in the 0.5-7 keV band using the same 3'' radius aperture and verified that below 0.9 keV and above 4 keV the source emission is indistinguishable from the background. The spectrum and  response files were created using the {\tt psextract} script in
CIAO. Since {\tt psextract} does not account for the PSF fraction when building up the effective area file, we multiplied the 0.5-2 keV and 2-10 keV fluxes as obtained from the spectral fit by a factor of 2 and 2.5, respectively, to recover the full aperture-corrected X-ray fluxes. We found consistent results either using spectral responses extracted from individual observations, or those obtained as an exposure-weighted mean over all individual responses. To double check the reliability of this procedure we also built spectral response files for one of the $Chandra$ exposures (ObsID=8594) using the ACIS-Extract software \citep{broos10} which allows proper construction of effective area files at any PSF fraction. Again, consistent results are found when using the ACIS-Extract responses.
We analyzed the X-ray spectrum with XSPEC v11.3.2 using the Cash statistic \citep{cash79} to estimate the best-fit parameters.
Errors are quoted at $1\sigma$ confidence level. We first fitted the data using a powerlaw spectrum modified by galactic absorption, which returns $\Gamma=-0.64^{+1.15}_{-0.73}$. We then used the {\tt plcabs} model \citep{yaqoob97}, which follows the propagation of X-ray photons within a uniform, spherical obscuring medium, accounting for both photoelectric absorption and Compton scattering. This model can be used in the case of
heavy absorption (up to $5\times10^{24}$ cm$^{-2}$) and up to rest-frame energies of $\sim 20$ keV. When fixing the photon index to $\Gamma=1.8$, we derived a Compton-thick column density of $N_H=1.4^{+0.9}_{-0.5}\times 10^{24}$ cm$^{-2}$ (see Fig.~1).  If we conservatively assume $\Gamma=1.0$, which is $\gtrsim4\sigma$ off the average intrinsic AGN value, we still obtain $N_H>10^{24}$ cm$^{-2}$. The measured absorption should be interpreted as a lower limit, since a cold reflection model ({\tt pexrav}), corresponding to $N_H\gtrsim10^{25}$ cm$^{-2}$, provides an equally good fit. \footnote{ When grouping to a minimum of 1 count per bin, the value of the C-statistic over the degrees of freedom is 54.7/59 and 55.2/60 for the {\tt plcabs} and {\tt pexrav} model, respectively.} 
Fitting the data with the recent MYTorus model \citep{mytorus}, which accounts for a toroidal distribution of the obscuring matter,
again returns $N_H>10^{24}$ cm$^{-2}$. No prominent iron K$\alpha$ line is observed at 1.1 keV (i.e., 6.4 keV rest-frame), but this is not in conflict with the Compton-thick scenario. Indeed, the $observed$ EW   scales with $(1+z)^{-1}$, so that EW$_{rest}\sim$1-2 keV, as is typical of Compton-thick AGN, would translate into EW$_{obs}\sim$170-340 eV. We verified that such a weak line can be easily accomodated in the fit and that only loose upper limits can be derived for the equivalent width (EW$_{rest}<4.3$ keV at 90\% c.l.). The aperture-corrected X-ray fluxes, as extrapolated from the X-ray fit, are $f_{0.5-2}=4.2\pm1.5\times 10^{-17}$ erg cm$^{-2}$ s$^{-1}$ and $f_{2-10}=6.8\pm2.4\times10^{-16}$ erg cm$^{-2}$ s$^{-1}$. The intrinsic (de-absorbed), rest-frame 2-10 keV luminosity is $\sim 2.5\times 10^{44}$ erg s$^{-1}$, which places XID403 at the low end of the X-ray luminosity range for type-2 quasars. Admittedly,  the uncertainties in the geometry of the obscuring (and reprocessing) material 
might substantially affect the derivation of the intrinsic luminosity. However, we note that, if the spectrum were produced by pure reflection and a typical reflection efficiency of $\sim 2\%$ is assumed \citep{gch07}, the intrinsic luminosity would be even higher.

XID403 was detected in the 1 Ms CDFS catalog by \citet[][XID=618]{giacconi02}, with 
$f_{0.5-2}=1.6\pm0.5\times 10^{-16}$ erg cm$^{-2}$ s$^{-1}$ and $f_{2-10}<7.6\times10^{-16}$ erg cm$^{-2}$ s$^{-1}$. The $\sim4$ times larger soft X-ray flux is likely due to contamination from high background fluctuations over the larger (8'' radius) extraction region adopted in that catalog. We checked the photometry of the
2000, 2007 and 2010 periods separately using a smaller 3'' radius: no significant source variability is detected in any band. 
XID403 was below the detection thresholds of the 1 Ms CDFS catalog by \citet{alex03} and 2Ms CDFS catalogs by \citet{luo08}. 
%A rough spectral analysis for XID=618 was presented in \citet{tozzi06}, which returned  $N_H=7.2 ^{+5.2}_{-1.6}\times10^{23}$ for $\Gamma$ fixed to 1.8, in agreement 
%within less than 1$\sigma$ with our analysis.

\section{Discussion and Conclusions}

\subsection{SED fitting}

We searched for additional indicators of heavy obscuration by considering data at other wavelengths.
We first investigated the observed (i.e., not corrected for absorption), rest-frame 2-10 keV to $6\mu m$  luminosity diagnostic ratio (X/IR, e.g., \citealt{alex08}). We derived the X-ray luminosity from the spectral fit and used the $\sim4\mu m$ rest-frame luminosity (derived from the Spitzer/MIPS datapoint at 24$\mu$m) as a proxy
for the $6\mu m$ luminosity. The X/IR ratio of $\sim 5\times10^{-3}$ would place XID403 in the region populated by Compton-thick AGN \citep{alex08}.
When considering the $F(24\mu m)/F(R)$ vs $R-K$ color diagram elaborated by \citet{fiore09}, XID403 would fall in ``cell E'' , where a significant fraction ($\sim25-30\%$)
of galaxies is found to host a heavily obscured, candidate Compton-thick AGN.

We then built the Spectral Energy Distribution (SED) from the X-rays to the radio regime (see Fig.2). 
As already shown by \citet{coppin09}, the radio to FIR emission of this source is dominated by dusty star formation, at a rate of $\sim$1000 $M_{\odot}$ yr$^{-1}$. We note that X-ray binaries associated to such a high SFR would produce 
$L^{obs}_{2-10}\sim2-5\times 10^{42}$ erg s$^{-1}$ (\citealt{ranalli03,lehmer10}), similar to the $observed$ value. However, X-ray binaries have on average much softer spectra ($\Gamma\gtrsim1.5$, \citealt{rm06}) than observed ($\Gamma=-0.64$). If  absorption is invoked to explain such a spectral hardness, then intrinsic luminosities of $\approx10^{44}$ erg s$^{-1}$ are derived, which are incompatible with X-ray binary emission.

A reasonable match with the MIR to UV datapoints could be obtained with a reddened QSO template with $A_V\sim0.7-0.8$ (adopting the extinction curve of \citealt{gb07}). When converting the measured optical extinction into an equivalent hydrogen column density by applying the relation valid for the Milky Way ISM ($N_H\sim1.8\times10^{21}A_V$), we find $N_H\sim 1.3\times10^{21}$ cm$^{-2}$, which is three dex smaller than what is estimated from the X-ray spectral fit. A mismatch between the X-ray and the optically estimated column density, in the range of $\sim3-100$, is observed in local AGN, calling for a number of interpretations \citep[e.g., low dust-to-gas ratio;][]{maiolino01powder}.  The mismatch observed in XID403 is $\sim$1000, which would make this object extreme. 
%Furthermore, when using the dust mass as estimated from the FIR emission and assuming a dust-to-gas mass ratio typical of the ISM ($\sim 100$) , one would obtain a gas mass (a few $\times 10^{10}\;M_{\odot}$) consistent with that measured from the 
%CO(2-1) line \citep[][$\sim1.6\times 10^{10}\;M_{\odot}$]{coppin10}. Finally, since the radio to FIR emission can be all explained by star formation (see Fig.2), one can use 
%the gas mass estimated from the FIR emission and the limits to the extent of the radio emission ($\lesssim1.2"$ radius, corresponding to $\lessim8$ kpc at $z$=4.76) to get a lower limit to the column density to the nucleus of $N_H\gtrsim1.6\times 10^{10}\;M_{\odot}$/1.2 kpc$^2=8\times10^{21}$ cm${-2}$. The combination of the above arguments would then cast some doubts on interpreting the full UV to near-IR rest frame SED of XID403 as mostly due to a mildly reddened AGN and would point towards a much higher extinction to the nucleus.

In their SED analysis \citet{coppin09} suggest a different, stellar origin for the optical/UV rest frame emission. 
Following the parameterization used by \citet{vignali09} and \citet{pozzi10} for obscured AGN, we fitted the SED of XID403 with a stellar component, an AGN torus component, and a dusty starburst component. The dusty starburst is responsible for the bulk of the FIR to radio emission, while the AGN torus produces the entire emission at 24$\mu$m (4$\mu$m rest-frame). A galaxy template with $M_\star\sim1.2\times10^{11}\,M_{\odot}$, $A_V\sim1$ and a 
$\sim$1-Gyr-old constant star formation rate nicely fits the optical/UV$_{rest}$ data. However, a possible problem in interpreting the optical/UV$_{rest}$ emission as stellar light is that XID403 is pointlike in the deep HST/ACS images (CLASS\_STAR=0.99 in $i_{775}$ and $z_{850}$), which would imply a half-light radius of $<0.3$ kpc. Although very compact morphologies have been observed in a fraction of distant sub-mm galaxies \citep{ricciardelli10}, the pointlike nature of XID403, coupled to the presence of broad NV emission, might suggest that the optical/UV$_{rest}$ 
light has a nuclear origin. In particular, we could be looking at a fraction ($\sim10\%$) of nuclear radiation that leaks out without being absorbed or is scattered towards us and thus would be polarized.
If true, the effective extinction to the nucleus would be much higher than that estimated by fitting the whole MIR to UV emission with a reddened QSO template, being more in line with the large X-ray column density.
% In addition, XID403 is the only object among the V-dropouts spectroscopically confirmed at $z\sim5$ by \citet{vanzella09} that shows clear AGN signatures in its spectrum. In particular a broad
%NV emission line is observed. All the other objects do not show any AGN feature and at the same time have small stellarity parameters, suggesting they are all extended as expected for non-AGN galaxies. 
%Also, the average SEDs of sub-mm galaxies and of ULIRGs at $z\sim2$ are much redder in the optical/UV regime.
%This might hint at another interesting possibility: namely that the optical/UV rest-frame light is part ($\sim10\%$) of nuclear radiation that leaks without being absorbed or it is scattered nuclear light. 
This interpretation has been already proposed by \citet{polletta08} to explain the relatively blue optical/UV emission and broad line components of two sub-mm galaxies at $z\sim3.5$ hosting heavily obscured AGN, similar to XID403. Also, although the stellarity parameter is uncertain for the 
faint $K$-band detection, the decrease of CLASS\_STAR from 0.95 in $Y$ to 0.74 in $K$ may also suggest that the host galaxy contributes significantly only at $\lambda_{rest}>4000$\AA.
In Fig.2 we show a possible SED decomposition for XID403 obtained by adding an AGN torus component and a scattering component (corresponding to $\sim10\%$ of the AGN intrinsic UV emission\footnote{The intrinsic AGN UV emission is estimated by normalizing the QSO template of \citet{elvis94}  to the Spitzer/MIPS datapoint.}) to the SED of Arp220.  
In summary, the full SED analysis shows that XID403 is not a classic, type-2 QSO (i.e., a narrow-line, X-ray obscured AGN whose physical properties can be explained within the standard, geometry-based Unified Model; \citealt{norman02}), but points to a complex physical picture likely related to its active assembly phase.

\begin{figure}
\includegraphics[angle=0,scale=.43]{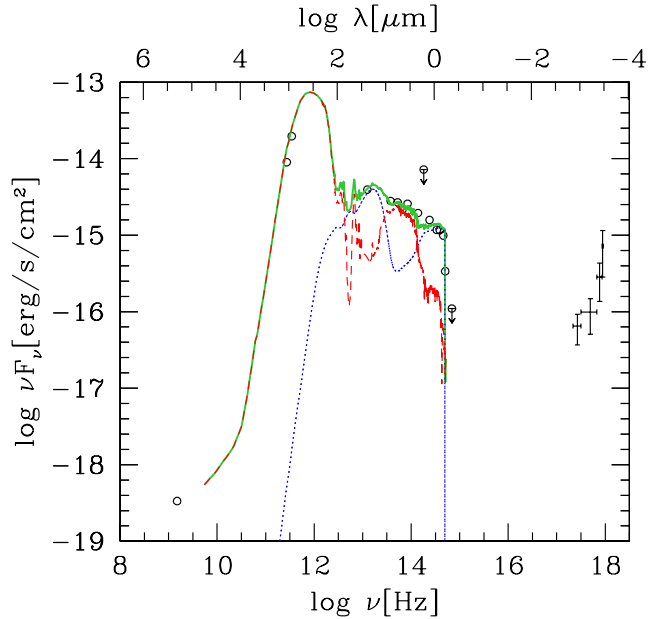}
\caption{Spectral energy distribution of XID403 (observed frame) and possible decomposition into a galaxy component (red dashed line) and
an AGN component (blue dotted line). The green solid line is the sum of the two. The template of Arp 220 shifted to $z=4.76$ is adopted for the galaxy emission. 
The sum of a torus and a scattered component is adopted for the AGN emission (see text).}
\end{figure}

\subsection{Black hole and stellar mass growth}

The IR emission from the AGN torus and the measured X-ray emission are used (see e.g., \citealt{vignali09,pozzi10}) to derive an AGN bolometric power of $7\times10^{45}$~erg~s$^{-1}$.
Assuming that the BH is radiating at the Eddington limit, as might be expected during these active BH and galaxy build-up phases, would imply 
$M_{BH}=5\times10^7\,M_{\odot}$. This in turn gives $M_{BH}/M_{\star} \sim 4\times10^{-4}$, which is a factor of 5 smaller than the local value. It would then seem that both BH and stellar mass are rapidly growing towards their final values, but the BH is lagging behind as seen in sub-mm galaxies at $z\sim2$ \citep{alex08submm} and expected by recent semi-analytic models of BH/galaxy formation \citep{lamastra10}. However, since the estimated BH mass is a lower limit (accretion might be sub-Eddington) and the stellar mass derived in the previous section might be an upper limit (the AGN likely contributes to the optical/UV$_{rest}$ light), this ratio might well be equal to the local value. 
%The relative rates of accretion to star formation, $\dot{M}_{Edd}/SFR = 1.1/1000=1.1\times10^{-3}$ \footnote{A standard radiative efficiency of $\epsilon=0.1$ is assumed
%to compute $M_{Edd}=L_{Edd}/(\epsilon c^2)$.}, may also suggest that the black hole and stellar mass keep sitting on the local relation while growing.

\subsection{Expectations for high-z Compton-thick AGN}

While the space density of luminous, unobscured and moderately obscured QSOs declines exponentially at $z\gtrsim3$ \citep[e.g.,][]{brusa09,civano11}, the behaviour of heavily obscured objects has still to be properly determined. Semi-analytic models of BH/galaxy evolution linking the obscuration on nuclear scales to the gas availability in the host galaxy \citep[e.g.,][]{menci08}, would predict an increasing abundance of obscured AGN towards high redshifts, and some observational evidence
of this trend has been reported (\citealt{tuv09} and references therein).

We considered the number of Compton-thick AGN as expected from the synthesis model by \citet{gch07}. XID403 is detected
with a 2-10 keV flux $\sim2$ times larger than the detection limit at its position. The mean limiting flux over the 160 arcmin$^2$ GOODS-S area is $f_{2-10}=1.5\times10^{-16}$ erg cm$^{-2}$ s$^{-1}$. Using the \citet{gch07} model,
one would expect from 0.06 to 0.6 Compton-thick AGN in the range of $4.7<z<5.7$ with $f_{2-10}>3\times10^{-16}$ erg cm$^{-2}$ s$^{-1}$ in GOODS-S, depending on
whether their space density undergoes the same high-z decline as observed for less obscured QSOs or stays nearly constant. 
Clearly, any firm conclusion is prevented by the low statistics. However, the mere presence of a Compton-thick AGN at $z>4$ in such a small area (and this could be a lower limit since we did not investigate the whole X-ray source catalog) might suggest that the space
density of Compton-thick AGN is not rapidly declining towards high redshifts. This shows that the detection of even a small number of heavily obscured AGN at $z>4$ in ultra-deep X-ray surveys would have a strong leverage on our understanding of early BH evolution.

X-ray spectral analysis is the only unambiguous way to determine whether an AGN is shrouded by Compton-thick matter. Observations at energies above 10 keV are an
obvious way to identify Compton-thick AGN, but the current high-energy instrumentation and that 
foreseen in the near future ($NuSTAR$, $Astro-H$) will not allow sampling objects beyond $z\sim1-1.5$.
Below 10 keV,  sensitive observations with limiting fluxes of $\approx10^{-17}$ erg cm$^{-2}$ s$^{-1}$ over wide sky areas,
such as those from the proposed missions $IXO$ \citep{white10} and $WFXT$ \citep{murray10}, would be required. The only concrete way to detect and unambiguously recognize high-z Compton-thick AGN in the near future is through even deeper observations with $Chandra$.

\acknowledgments
We acknowledge the following supporting agencies and grants:
Italian Space Agency (ASI) under the ASI-INAF contracts I/009/10/0 and I/088/06/0 (RG, CV, AC);
NASA through CXC grant SP1-12007A and ADP grant NNX10AC99G (WNB, YQX, BL);
$Chandra$ archival grant SP89004X (AP, RG);  DFG cluster of excellence Origin and Structure of 
the Universe (PR). We thank the referee  for a constructive report and N. Miller and M. Mignoli for useful discussions.
RG gratefully acknowledges hospitality by the Johns Hopkins University where 
this work started.

\end{document}